\documentclass[11pt]{article}
\usepackage{jcappub}
\usepackage{array}
\usepackage{aas_macros}
\usepackage{graphicx}
\usepackage{epsfig}
\usepackage{subfigure}
\usepackage{mathrsfs}
\usepackage[normalem]{ulem}

\newcommand{\be}{\begin{equation}}
\newcommand{\ee}{\end{equation}}
\newcommand{\mpc}{\, {\rm Mpc}}

\newcommand{\hmpc}{\, h^{-1} \mpc}

\newcommand{\FastPM}{\textsf{FastPM}}

\title{A gradient based method for modeling baryons 
and matter in halos of fast simulations}  
\author[a,b]{Biwei Dai,}
\author[b]{Yu Feng,}
\author[b,c,d]{Uro\v s Seljak}

\affiliation[a]{Department of Physics, Peking University, Beijing, 100871, China}
\affiliation[b]{Berkeley Center for Cosmological Physics and Department of Physics, University of California, Berkeley, CA 94720, USA}
\affiliation[c]{Department of Astronomy, University of California, Berkeley, CA 94720, USA}
\affiliation[d]{Lawrence Berkeley National Lab, 1 Cyclotron Road, Berkeley, CA 94720, USA}

\emailAdd{biwei@berkeley.edu}

\abstract{Fast N-body PM simulations with a small number of time steps such as FastPM or COLA
have been remarkably successful in modeling the galaxy statistics, but their lack of 
small scale force resolution and long time steps cannot give accurate halo matter 
profiles or matter power spectrum. 
High resolution N-body simulations can improve on this, but lack baryonic effects, which 
can only be properly included in hydro simulations. 
Here we present a scheme to calibrate the fast simulations to mimic the precision of the
hydrodynamic simulations or high resolution N-body simulations. The scheme is based on 
a gradient descent of either effective gravitational potential, which mimics the short range force, or of effective 
enthalpy, which mimics gas hydrodynamics and feedback. The scheme is fast and 
differentiable, and can be incorporated as a post-processing step into any simulation. It 
gives very good results for the matter power spectrum for several of the baryonic 
feedback and dark matter simulations, and also gives improved dark matter halo profiles. The scheme is even able to find the large 
subhalos, and increase the correlation coefficient between the fast simulations and the 
high resolution N-body or hydro simulations. 
It can also be used to add baryonic effects to the high resolution N-body simulations. 
While the method has free parameters that 
can be calibrated on various simulations, they can also be 
viewed as astrophysical nuisance parameters describing baryonic effects 
that can be marginalized over during the data analysis. In this view these
parameters can be viewed as an efficient parametrization of baryonic effects.}

\begin{document}
\maketitle
\flushbottom

\section{Introduction}

Extracting accurate cosmological information from the current and future sky surveys requires high precision simulations. Computer simulations with quasi N-body numerical schemes provide an alternative to the full N-body or hydro simulations for creating fast realizations of the 
large scale structure (LSS), but lack resolution on small scales. Methods in this family includes \FastPM{} \citep{feng2016a}, COLA \citep{tassev2013a}, and subcycle TreePM \citep{sunayama2016a}. 

Although the halo catalog from the quasi N-body simulations are well correlated to a true N-body simulation of the same initial condition \citep{feng2016a}, the dark matter density distribution is less accurate, since it requires not just the center of mass of a 
dark matter halo but also its density profile. For example, a 10 step \FastPM{} N-body simulation misses about $15\% - 20\%$ of power at the Nyquist frequency of the PM grid 
\citep{feng2016a, tassev2013a}. Accurate modeling of the dark matter requires more time steps. For example, Izard et al. \citep{Izard2016a} proposed ICE-COLA, in which the code parameters are optimized to achieve a matter power spectrum within 1 percent for $k \lesssim 1 \mathrm{h Mpc^{-1}}$, at the cost of using 40 time steps and a force mesh that is 3 times smaller than the particle mean separation. This leads to about an order of magnitude
higher cost than a 10 step simulation without a higher force mesh resolution. 

The suppression in the power spectrum is due to the inability of these methods to resolve the internal structures of halos. With such large time steps, the halo is not fully virialized, resulting in a much shallower density profile than the NFW profile \citep{navarro1996a,navarro1997a}. This severely undermines the application of quasi N-body simulations to the applications where modeling underlying dark matter fields are crucial, including SZ effects and weak lensing. 

On the other hand, even the most accurate N-body simulation cannot model the effect of baryons without introducing the costly and complex numerical schemes to model the hydrodynamics, cooling, star-formation and AGN feedback. The effects of baryon on the density profile are also important. Gravity leads to collapse, but baryons, due to the gas pressure, resists the collapse. In addition, the feedbacks from AGN and supernovae can 
transport a large amount of gas to the outskirts of the halos. This can even lead to an expansion of the dark matter halo and the reduction of the halo mass \citep{duffy2010a, teyssier2011a, mccarthy2011a, velliscig2014a}. It has been shown that these effects can reduce the matter power spectrum by more than ten percent for $1 \lesssim k \lesssim 30 \mathrm{hMpc^{-1}}$ \citep{vandaalen2011a,vogelsberger2014a}. 

In this paper we introduce a simple numerical scheme that allows an arbitrary calibration of the dark matter density field against high resolution N-body simulations and hydrodynamical simulations. Our method is based on motion of particles along the gradient direction of a scalar field that is generated from the existing density field. The method is simple in the sense that the Jacobian of the method has a very simple form, and can be easily embedded into a parameter inference framework. Our method is essentially a data driven learning and 
can be viewed as a form of machine learning (ML): rather 
than having a full dynamical model for the matter density distribution, we train the 
low resolution simulation to reproduce the results of the more expensive 
high resolution simulations. Training is performed on the comparison between the low and 
high resolution simulations. However, our specific approach is gradient based rather
than using standard ML techniques, which we argue offers several advantages, primarily low 
dimensionality of parameter space to be optimized against. 

The plan of the paper is as following: we first describe the mathematics and motivation of the numerical schemes in Section \ref{sec:models}.  In Section \ref{sec:calibration} we show the performance of the gradient based schemes in an example, with the emphasizes on matter power spectra, halo profiles and sub-halo statistics. We present a recommendation of the parameter choices in Section \ref{sec:parameter}. Finally we conclude in Section \ref{sec:conclusion}.

\section{Gradient based learning: theory and motivation}
\label{sec:models}

In this section we introduce the gradient based method and derive several forms 
depending on the problem we wish to solve. The basic goal is that we would like 
to have a scheme that mimics the physics that is missing in low resolution 
simulations. This can be either short range force in the case of gravity, or 
hydrodynamic force effects in the case of hydrodynamic simulations. 
Our main idea is that instead of performing a full simulation we can partially 
account for these effects as a post-processing step, which will perform the 
effective missing short range force operation only once. Since force is a 
gradient of some effective potential this leads to the idea of descending along 
its gradient. The parameters controlling the model can be learned from the 
high resolution simulations themselves, or more 
precisely by comparing high and low resolution simulations. This is a form of 
data driven learning, but since it is based on some notion of missing physics 
it can be parametrized with relatively few free parameters. 
We describe next several versions of this idea. 

\subsection{Descent along the gradient of gravitational potential: Particle-Particle 
interaction}
\label{sec:PGD}

Without the short range particle-particle force, quasi-nbody schemes such as \FastPM \cite{feng2016a} do not resolve the structures with scales smaller than the mesh resolution, resulting in halo profiles that are shallower than their full N-body counterparts. 
A straightforward way to improve this would be adding a short range Particle-Particle (PP) force in every step during the simulation, which would however also induce a large additional computational cost. 
However, the effect of this short range PP force is to aid the collapse of halos, which is a radial motion of the particles. We thus propose a simple model in which we move the particles along a short range PP force after the simulation has finished. The displacement bypasses the momentum of the particles, corresponding to the case of a fluid with infinite viscosity. 
The direction of the short range PP force points towards the potential minimum. Therefore, displacing particles along this direction can sharpen the profile of halos. 

A tree can be built to accelerate the calculation summation of short range PP force, 
\be
\mathbf{F_{\mathsf{PP}}} = \sum_{r < r_{max}} \frac{Gm_1m_2}{(r+\epsilon)^3} \cdot \mathbf{r}
\ee
where $r_{max}$ determines the largest separation of particle pairs to calculate the force, and $\epsilon$ is the force softening. The KDTree implementation in kdcount is improved to compute the short range force \citep{kdcount}. 

The displacement is proportional to the force and given by
\be
\label{eq:PGDPP}
\mathbf{S} = (\alpha_{\mathsf{PP}}/H_0^2)\ \mathbf{F_{\mathsf{PP}}}
\ee
where $\alpha_{PP}$ is a free parameter in our model and $H_0$ is the Hubble parameter induced here to make $\alpha_{\mathsf{PP}}$ dimensionless. 

We can vary the 3 parameters ($\alpha_{PP}$, $\epsilon$, and $r_{max}$) to match the power spectrum against N-body and hydrodynamical simulations 
or against averaged halo profiles. This is discussed further in section \ref{sec:method}. 

\subsubsection{Descent along the gradient of gravitational potential: Particle-Mesh interaction}
Particle-Mesh (PM) force is significantly cheaper to compute than a PP force. We therefore consider a faster scheme where the short range force is instead computed with a Fourier space particle mesh solver. The full gravitational potential of a particle mesh solver is given by
\be
\phi = 4 \pi G \bar \rho \nabla^{-2} \delta,
\ee
where the force is given by the gradient of the potential.

\begin{figure}
\includegraphics[width=\textwidth]{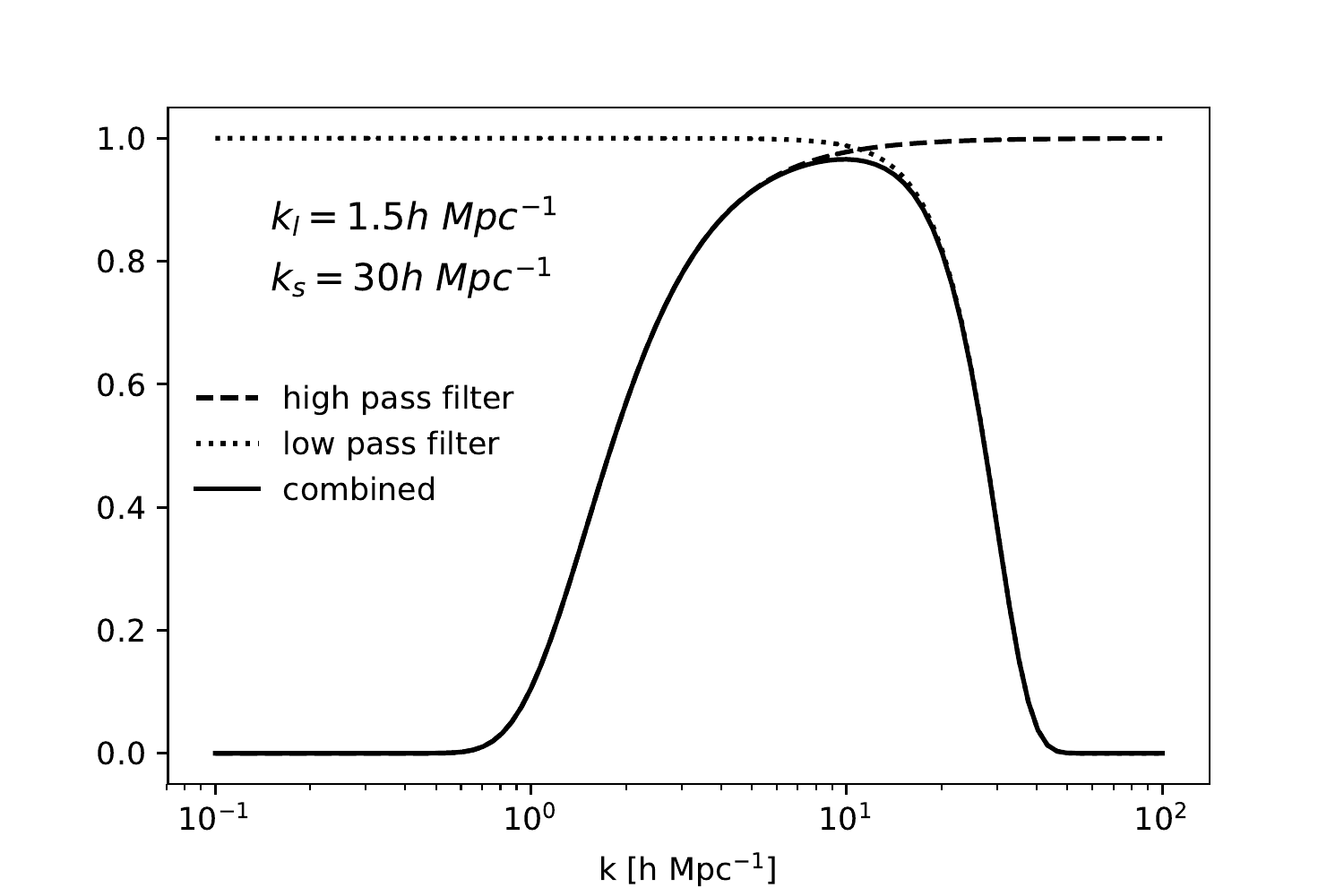}
\caption{The filter as a function of k.}
\label{fig:filter}
\end{figure}

Drifting the particles along the gradient of the gravitational potential will act as an additional time step, increasing the large scale growth. We can eliminate the large scale component of of the potential with a high pass filter $\widehat{O_1}$,
\be
\mathbf{\hat{O}_l}(k) = \exp{(-\frac{k_l^2}{k^2})}
\ee
where $k_{l}$ is the long range scale parameter. The spirit of the high pass filter is similar to the $r_{max}$ parameter in our PP model: this filter removes the long range force by damping the potential modes with scales larger than $k_{l}$. On small scale, to reduce the numerical effect induced by the mesh resolutions, we introduce another low pass filter
\be
\mathbf{\hat{O}_s}(k) = \exp{(-\frac{k^4}{k_s^4})} 
\ee
where $k_{s}$ is the short range scale parameter. The low pass filter has a similar effect as force softening $\epsilon$, although the cut-off is slightly sharper. We show an example of the filter in Figure \ref{fig:filter}.

We compute the displacement with the filtered potential:
\begin{align}
\mathbf{S} &= (\alpha_{\mathsf{PM}}/H_0^2)\ \mathbf{F_{\mathsf{PM}}}\\
&= (\alpha_{\mathsf{PM}}/H_0^2)\ \mathbf{\nabla}\mathbf{\hat{O}_l\hat{O}_s}\phi\\
&= (4 \pi G \bar \rho\alpha_{\mathsf{PM}}/H_0^2)\ \mathbf{\nabla\hat{O}_l\hat{O}_s} \nabla^{-2} \delta
\label{eq:PGDPM}
\end{align}

\begin{figure}
\includegraphics[width=\textwidth]{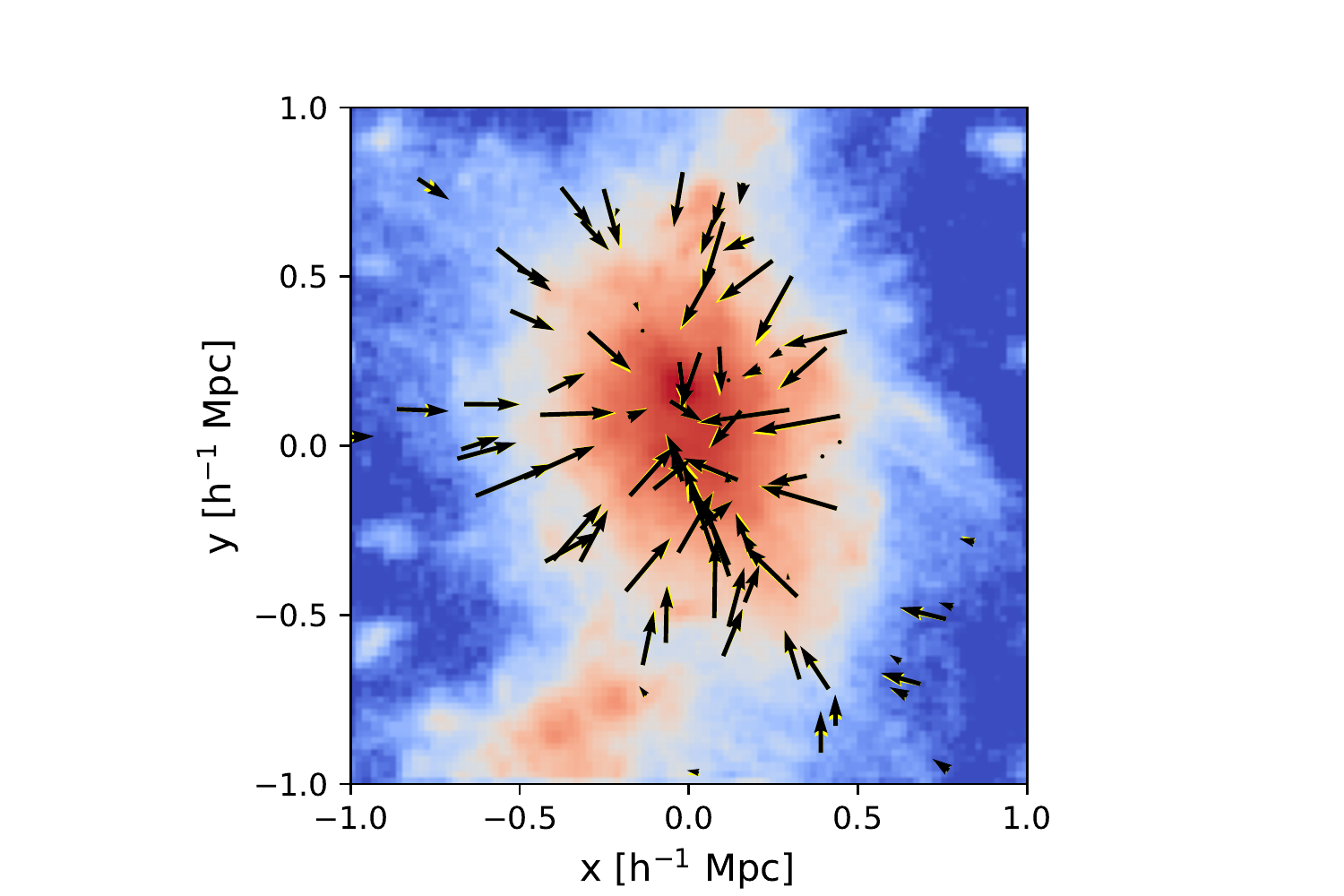}
\caption{A visualization of the descent along gradient of gravitational potential. The black arrows show the displacements calculated from particle-mesh interaction (Equation \ref{eq:PGDPM}), while the yellow color present the displacements from particle-particle interaction (Equation \ref{eq:PGDPP}). The yellow arrows mostly overlap with black ones and can hardly been seen. Here we only show the displacements of a sample of particles in the halo.}
\label{fig:pp-vs-pm}
\end{figure}
In Figure \ref{fig:pp-vs-pm}, we show that Equation \ref{eq:PGDPM} achieves a similar effect comparing Equation \ref{eq:PGDPP}, but the former one is much faster, suggesting that a full resolution PP force solver is unnecessary.  
We will only show the results from the PM method in the rest of this paper. We name this scheme as the Potential Gradient Descent model (hereafter PGD).

\begin{figure}[ht]
\includegraphics[width=\textwidth]{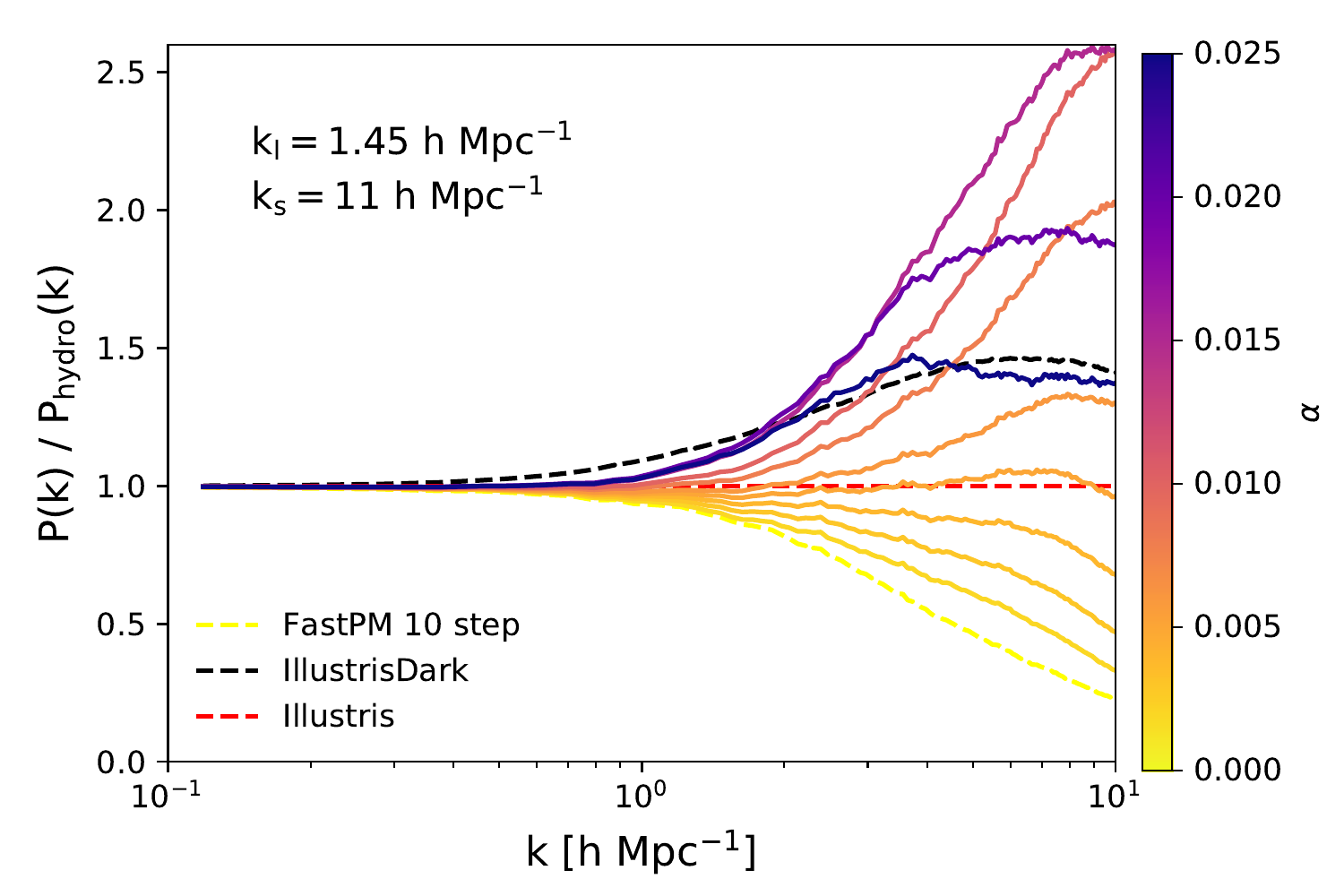}
\caption{The power spectra of 10 step \FastPM{} simulation using PGD models with different $\alpha$. At $\alpha=0.025$ we can match the effects of pure dark matter 
high resolution simulation (IllustrisDark) and at $\alpha=0.005$ that 
of the hydrodynamic simulation (Illustris). }
\label{fig:illustris-power-varyalpha}
\end{figure}

In figure \ref{fig:illustris-power-varyalpha} we show how the parameters influence the matter power spectrum. Here we focus on the parameter $\alpha$ in PGD model fixing the parameters $k_l$ and $k_s$. As expected, as $\alpha$ increases, the halo profiles are sharpened and the small scale power spectrum increases. When $\alpha \gtrsim 0.015$, the particles begin passing the halo centers and increasing $\alpha$ will smooth the density field, making the small scale power spectrum decrease. The maximum power enhancement is achieved at $\alpha=0.015$ (depending on $k_l$ and $k_s$). 
We see from the figure that varying $\alpha$ can match both the effects of pure dark matter 
high resolution simulation (IllustrisDark) and of the hydrodynamic simulation (Illustris). 

The PGD model is similar to the Augmented Lagrangian Perturbation Theory (ALPT) \citep{kitaura2013a}. However, ALPT assumed the analytical spherical collapse model and determined the small scale displacement using local density, while our method is based on a PM solver and solves a modified gravitational potential. Our method has more freedom, with 3 free parameters rather than 1, which makes it more effective in terms of matching to high resolution simulations (see Section \ref{sec:calibration}).

\subsection{Descent along the gradient of enthalpy}
\label{sec:EGD}

The effect of baryons on the power spectrum can be viewed from two aspects. On scales $k \approx [0.3,30] \mathsf{hMpc^{-1}}$, pressure, stellar and AGN feedback smooths the density field and reduce the power. On even smaller scales, the power is increased because of the cooling \citep[e.g.,][]{vandaalen2011a, vogelsberger2014a}. We will focus on $k < 10 \mathsf{hMpc^{-1}}$, where the effect of cooling is sub-dominant. The two remaining effects are modeling pressure and feedback, both of which transfer matter from the 
high density regions to the low density regions. 

Motivated by the hydro-PM (HPM) simulation \citep{gnedin1998a}, we define a pressure-like potential (specific enthalpy). We assume that to the first order, the distribution of baryon, dark matter and the total matter are the same. The density field is first smoothed with a Gaussian kernel
\be
\mathbf{\hat{O}_{J}}(k) = \exp(-\frac{(kr_{\mathsf{J}})^2}{2})
\ee
where $r_{J}$ is the smoothing scale and we set it to be $0.1 \mathsf{h^{-1}Mpc}$, of 
the order of the Jeans scale. We also assume a power law equation of state, \citep[e.g.][]{gnedin1998a},
\be
T(\delta) = T_0 (1+\delta_b)^{\gamma-1}
\ee
where $T_0$ is a constant and we set it to be the characteristic temperature of IGM ($10^4 \mathsf{K}$). The HPM method typically takes $\gamma = 1.4 \sim 1.6$ for the low density IGM. In our case, the equation of state is an effective one due to the feedback of star formation and AGN feedback, and therefore we expect $\gamma$ to be a free parameter. The pressure $p$ can be easily calculated once the temperature and density are known: 
\begin{align}
p(\delta) &= n_b k_B T(\delta)\\
&= \frac{\bar{\rho_b }k_B T_0}{\mu}(1+\delta)^{\gamma}
\end{align}
where $\bar{\rho_b}$ is the averaged baryon density, $k_B$ is the Boltzmann constant, and $\mu$ is the averaged atomic mass of the gas which we set to be the Hydrogen atomic mass. Now we introduce the specific enthalpy $\mathcal{H}$:
\be
\mathcal{H} = \frac{P(\rho)}{\rho}+\int^{\rho}_1\frac{P(\rho')}{\rho'}\frac{d\rho'}{\rho'}
\ee
The displacement is given by:
\begin{align}
\mathbf{S_b} &= - \frac{\beta}{H_0^2} \mathbf{\nabla} \mathcal{H}(\delta)\\
&= - \frac{\beta}{H_0^2} \frac{k_BT_0}{\mu} \frac{\gamma}{\gamma-1} \mathbf{\nabla} [\mathbf{\hat{O}_{J}}(1+\delta)]^{\gamma-1}
\end{align}
where $\beta$ is the scale factor and a free parameter in the model, $H_0$ is the Hubble parameter to make $\beta$ dimensionless. $T_0$ and $\mu$ is degenerate with the parameter $\beta$, so here we only assign them with the correct order of magnitudes and do not attempt to model their accurate values. It is clear in the above equation that the effective equation of state $\gamma$ essentially determines how the displacement depends on the density field, so we expect that changing this parameter will be able to model the halo mass dependence of AGN feedback. We note that \cite{gnedin1998a} applied a pressure uniformly to all particles, while in our model each particle has the probability of $\frac{\Omega_b}{\Omega_m}$ to be identified as a baryon and hence be displaced. The rest of the dark matter particles are not displaced in this model. 
We found that this schemes performs best for our applications. We will refer to this model as Enthalpy Gradient Descent (EGD) in the rest of this paper.

\subsection{Radial flows towards the halo center}

Here we describe an alternative approach that is not gradient based. 
The models presented above move particles without any knowledge where the halos are located or which halo a given particle belongs to. Instead, we asserted that the centers of halos have minimum potential and maximum density (therefore maximum pressure), even if they are not sufficiently prominent to be identified by any halo finder. If a particle is located within a halo, it is likely to be moving along the radial direction; and if it is outside the halo, the gradient will be small and so is the displacement. Alternatively, we can directly solve for the radial displacement of particles toward halo centers such that the the spherical averaged radial density profile matches the halos found in hydrodynamical simulations. 

In such a scheme, 
first the friends-of-friends (FOF) halo finder is performed to find all the halos that we wish to calibrate the profiles in both the reference and the quasi-nbody simulations. Then the centers of halos (defined as the potential minimum)\footnote{Density maximum is noisy due to the small scale fluctuations.} are found, and particles are assigned labels according to their host halo. We assign unlabeled particles to the nearest halo. 

The halo profiles and baryonic feedback are halo mass dependent. We can divide the halos into different mass bins and measure the averaged enclosed mass as a function of radius $M_i(r)$, where $i$ denotes different halo mass bins. Then for each halo mass bin, we find the abundance matched halos in the hydrodynamical simulation and measure the corresponding enclosed mass $M_{\mathsf{ref},i}(r)$. $M_{\mathsf{ref},i}(r)$ is a monotonic increasing function so we can write its inverse function
\be
r_{\mathsf{ref},i}(M) =  M_{\mathsf{ref},i}^{-1}(r)
\ee
We define the radial displacement $D(r)$ as a function of radius:
\be
\label{eq:RadialD}
D(r) = r_{\mathsf{ref},i}[M_i(r)] - r
\ee
The definition above ensures the enclosed mass after calibration is the same with the reference simulation, so that the halo profile, which corresponding to the derivative of $M_i(r)$, also matches. Equation \ref{eq:RadialD} also makes sure that the spherical shells of a halo does not cross with each other during the calibration.

A naive displacement may leave gaps at the edge of halos. We therefore apply a smooth truncation to the displacement according to a characteristic radius $r_{500}=(\frac{3M_{fof}}{2000\pi})$. If a particle is located outside this radius, we suppress the displacement by a factor of $\exp(1-\frac{r}{r_{500}})$. 

The full formula for the displacement is
\be
\mathbf{S} = 
\begin{cases}
D(r) \mathbf{e_r} &r \le r_{500}\\
D(r) \exp(1-\frac{r}{r_{500}})\mathbf{e_r} &r > r_{500}
\end{cases}
\ee
where $\mathbf{e_r}$ is the unit vector of the radial direction. 

This method is the most direct way to manipulate the halo profiles, but it is not gradient 
based, or based on any other physical considerations. Since this model requires lots of computations including running FOF halo finder, finding the potential minimum as the halo centers and measuring the halo profiles, it is much slower than the potential and enthalpy gradient descent models introduced above. Taking a gradient of the final density with respect to some initial modes (the Jacobian) is also problematic, as the process involves non-differentiable procedures, such as peak finding, binning, and connecting friends-of-friends halos. The method does not enhance the internal substructures of a halo because the radial displacement is not aware of any substructures. Because of all these reasons this method is less suitable for our purposes, but we have nevertheless developed it and present results below. This method will be referred to as Radial Flow (RF) model.

\section{Example application: matching a set of hydrodynamical and N-body simulations}
\label{sec:calibration}

As an example, we calibrate 3 ''inaccurate'' simulations against Illustris hydrodynamical simulations \citep{vogelsberger2014a, vogelsberger2014b,genel2014a}: \FastPM{} with 10 steps, \FastPM{} with 40 steps, and high resolution dark-matter-only simulation (IllustrisDark). These simulations use the same linear power spectrum and the same random seed with the hydrodynamical simulation. The Python version of FastPM was improved to perform the FastPM simulations used in this work \citep{fastpm-python}. 
The power spectra of these simulations at redshift 0 before calibration are shown in Figure \ref{fig:powerspectra}:
\begin{itemize}
\item The 10 step \FastPM{} lacks small scale power. Therefore we apply the potential gradient descent model to sharpen the halos and increase the power on small scales. 

\item  The high resolution N-body simulations is over-clustered at small scale due to the lack of feedback processes. Therefore we mimic the baryonic effects and lower the power using our enthalpy gradient descent model.

\item  \FastPM{} with 40 steps: on scales $0.5 \mathsf{hMpc^{-1}} \lesssim k \lesssim 5\mathsf{hMpc^{-1}}$ it has more power than the reference simulation because of the absence of baryonic feedback, while on scales $ k \gtrsim 5\mathsf{hMpc^{-1}}$, the power is reduced and is similar to 10 steps. Therefore, we apply both potential and enthalpy models to 40 steps \FastPM{}. 
\end{itemize}

We vary the free parameters in the models to fit the power spectra and find the best fit solutions with the maximum likelihood defined in Equation \ref{eq:likelihood}.

There are 3 free parameters in the potential gradient descent model: $\alpha_{\mathsf{PM}}$, $k_l$ and $k_s$, and 2 parameters in the enthalpy gradient descent model: $\gamma$ and $\beta$. We ask $\gamma > 1$ during our fitting, so that the matter are moving from high density regions to low density regions.

\subsection{Visual Inspection}

\begin{figure}[ht]
\includegraphics[width=\textwidth]{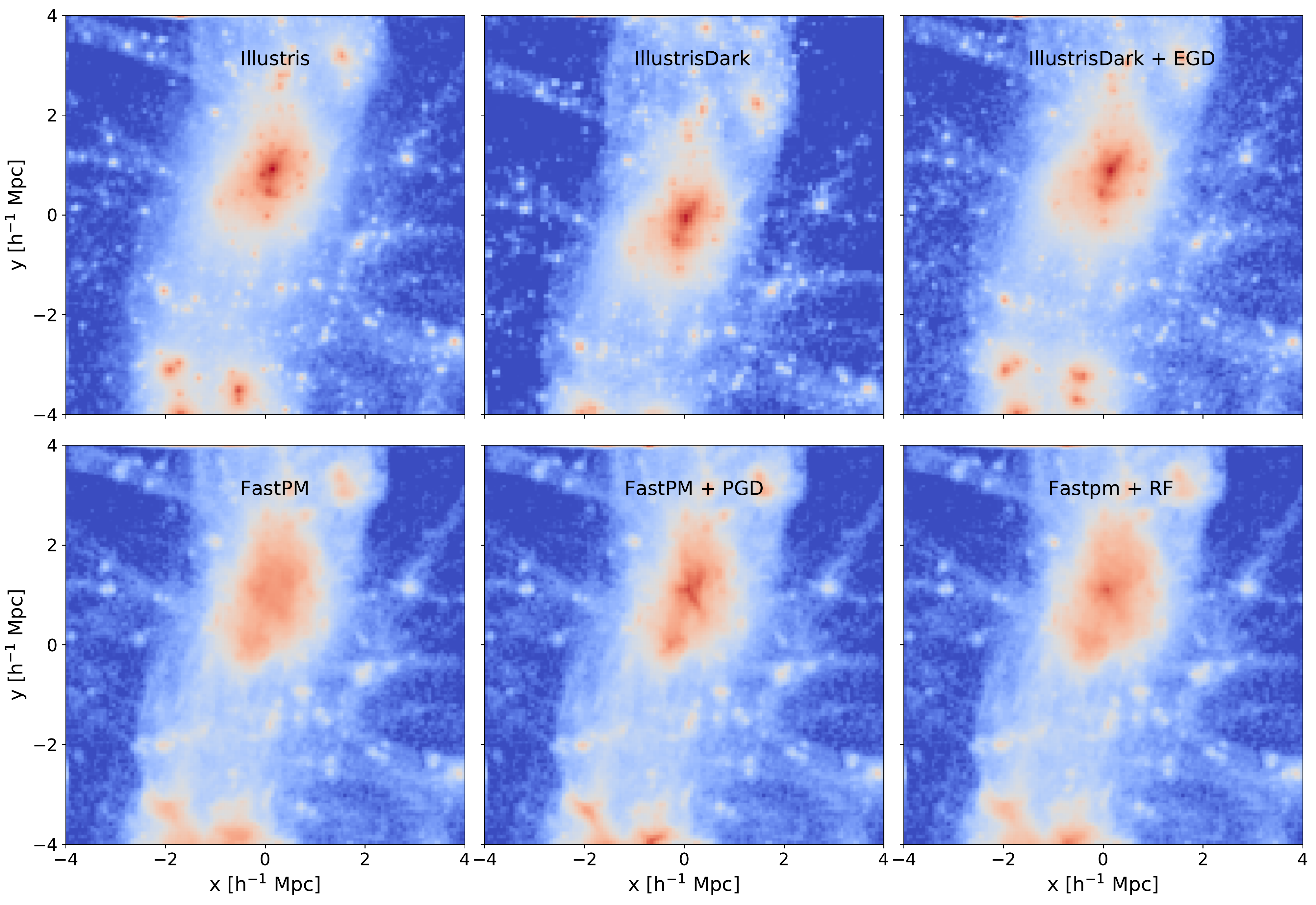}
\caption{The projection of the same halo in Illustris, IllustrisDark, and \FastPM, before and after applying our models. The top left panel is from Illustris-3, the top middle panel shows IllustrisDark, and the top right one is IllustrisDark after applying our enthalpy gradient descent model (see Section \ref{sec:EGD}). The enthalpy gradient descent model moves some particles from halo center to the outer region to simulate AGN feedback, making the outskirts of IllustrisDark closer to Illustris (see the upper right corner of each plots in the upper panel). The bottom panels show the halos in \FastPM{} simulation with 10 step, before and after applying our potential gradient descent model and radial flow model, respectively. The halo in \FastPM{} does not fully collapse and no density peak can be found. After applying our models, the peaks appear. The radial flow model produces a smooth density profile with correct spherical density profile, but the potential gradient descent model creates some of the substructures and looks closer to Illustris.}
\label{fig:illustris-halo-projection}
\end{figure}

Before quantitatively presenting the results of our calibration, we first show a visual impression of how our models modify the matter distribution in a single halo in Figure \ref{fig:illustris-halo-projection}.  Comparing to dark-matter-only simulation, AGN feedback in hydrodynamical simulation moves a large amount of gas from the center of halos to large radii \citep{duffy2010a, mccarthy2011a}. We can see this effect in Figure \ref{fig:illustris-halo-projection}, where the halo from Illustris appears fuzzier at the outskirts than IllustrisDark (the upper right corner, in the upper left panel and upper middle panel). Our enthalpy gradient descent procedure successfully models this effect by pushing particles away from the halo, producing a smoother density field. The projected density field (especially at large radii) looks closer to the hydrodynamical simulation after applying our model.

Figure \ref{fig:illustris-halo-projection} also shows that the inner profile of the halo in 10-step \FastPM{} is not cuspy enough. As discussed above, this is mostly due to the limited force resolution of particle mesh and the insufficient number of steps which limits the nonlinear effects. Both potential gradient descent model and radial flow model produce a cuspy center by moving the particles towards the center and contracting the halo. However, the density field is still relatively smooth after applying the RF model because the particles are moved isotropically, and no evident substructures can be found. The potential gradient descent model, on the other hand, is able to model some of the substructures. This is not surprising. Even though halos in Fastpm simulation have not fully collapsed, we expect that the seeds of these substructures remain in the density field, and the gradient descent can amplify these density fluctuations.

In the rest of this subsection we will present in detail how these models improve the matter power spectra, halo profiles and sub-halo statistics.

\subsection{Power Spectra of Matter}

\begin{figure}[ht]
\includegraphics[width=\textwidth]{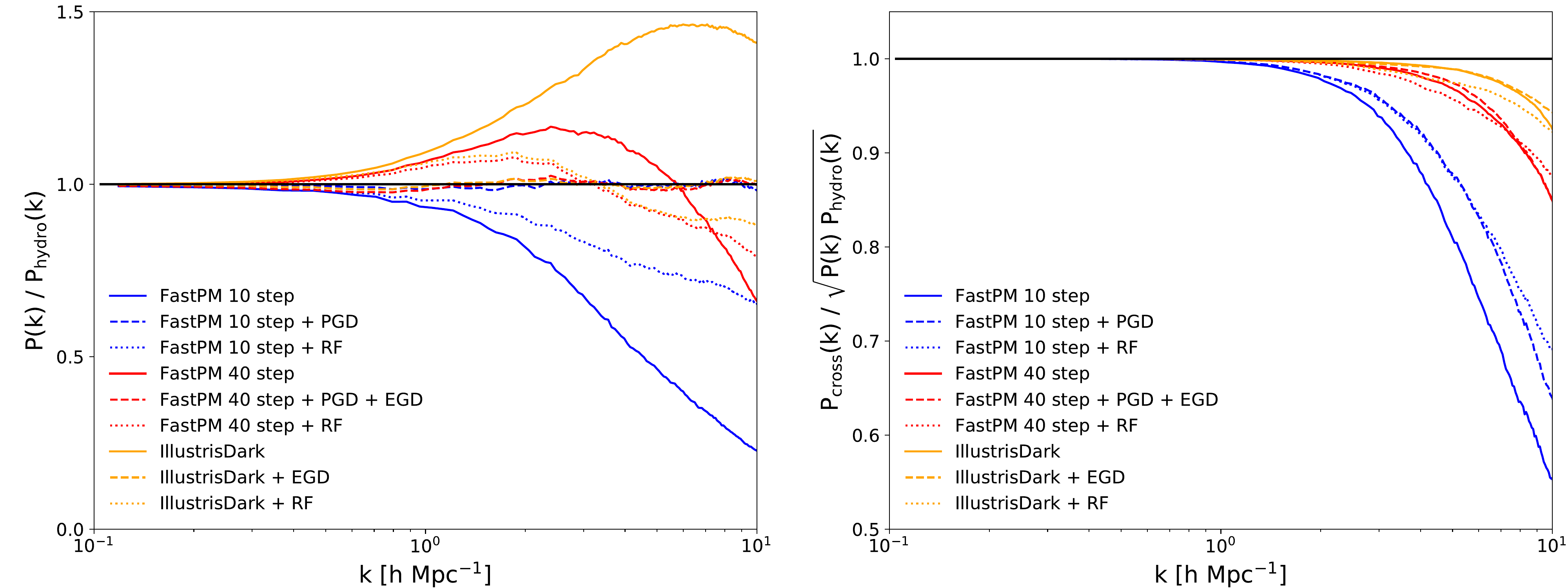}
\caption{The ratio of matter power spectra  (left panel) and cross correlation coefficients (right panel) of \FastPM{} with 10 steps, 40 steps, and dark-matter-only simulation, before and after using our models, compared to hydrodynamical simulations. 
The straight lines show results before calibration, while the dashed line and dotted line present the results after applying the models. After calibration, the deviations of matter power spectra compared to hydrodynamical simulation are within $5\%$. The cross correlation coefficients also improve after calibration. The power spectrum is calculated using Nbodykit \citep{nbodykit}.}
\label{fig:powerspectra}
\end{figure}

Figure \ref{fig:powerspectra} shows the matter power spectra and cross correlation coefficients of \FastPM{} and dark-matter-only simulations comparing to hydrodynamical simulations, before and after applying the model. For 10 step \FastPM{} and dark-matter-only simulation, the potential and enthalpy gradient descent models work fine, reducing the relative deviations of power spectra to within $5\%$.

We also observe that the cross correlation coefficient of 10 step \FastPM{} improves after calibration. We point out that the improvement is better than 11 step \FastPM{} simulation, although the computational cost is the same.


The RF model is not based on optimizing the power spectra, as it calibrates the halo profiles and improves the one halo term in the halo model, leading to better small scale power spectra. We notice that the improvement of cross correlation coefficients of PGD and RF are similar.

\subsection{Mass Profile of Halos}

\begin{figure}[ht]
\includegraphics[width=\textwidth]{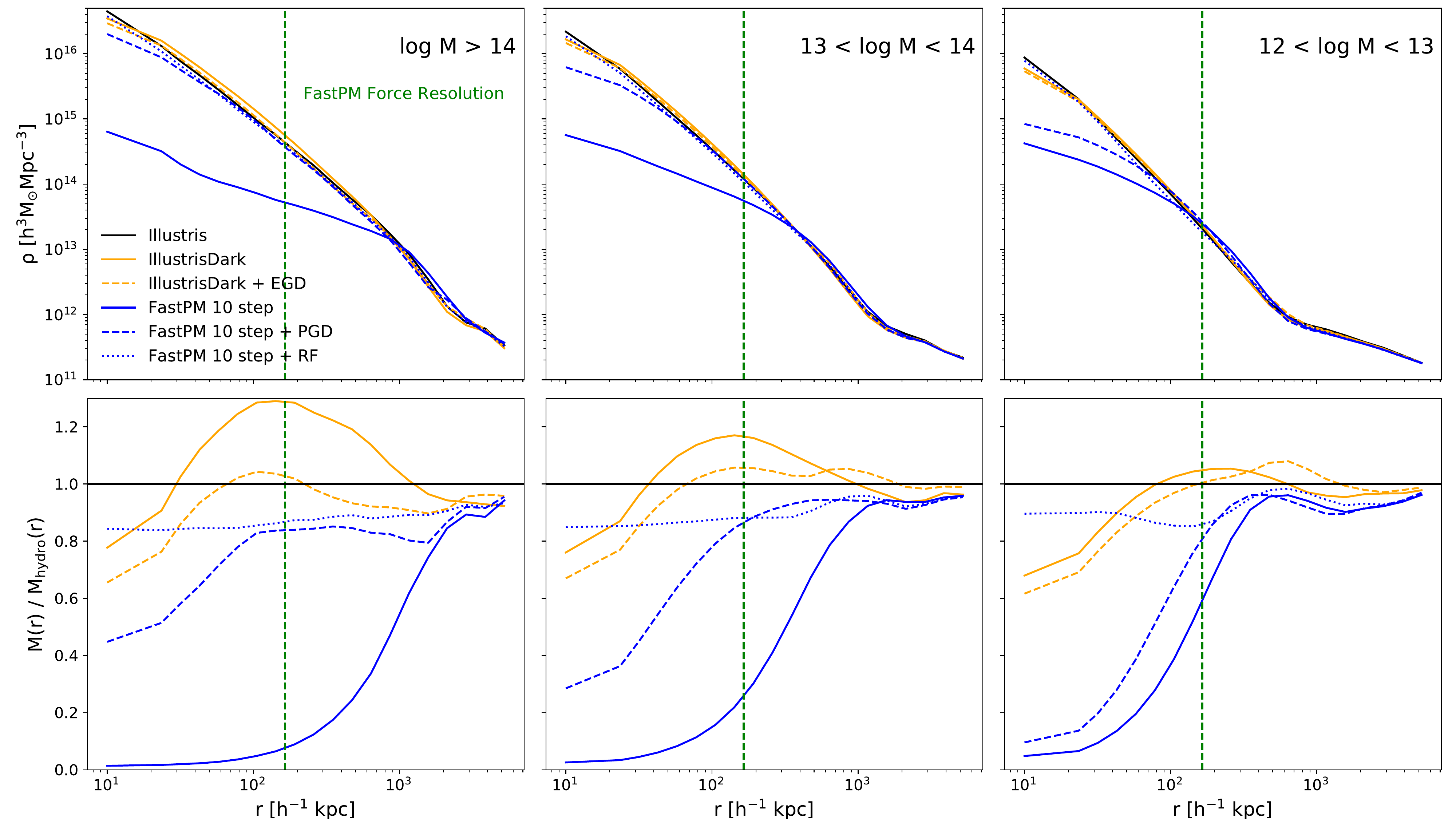}
\caption{The spherical averaged halo density profile (upper panels) and enclosed mass as a function of radius (lower panels), in different halo mass bins. Black color shows Illustris, yellow represents Illustris-Dark, and blue displays 10 step \FastPM{}. The solid line shows results before calibration, the dashed line and dotted present results after using different calibration models. The spherically averaged matter distribution gets closer to Illustris after calibration. The effect is stronger for larger halos.}
\label{fig:illustris-halo-profile}
\end{figure}

Figure \ref{fig:illustris-halo-profile} shows the calibrated \FastPM{} and Illustris-Dark halo profile against the reference Illustris-3 simulation. We see that both density profiles and matter profiles improve after calibration. This translates to a particularly large improvement of PGD for massive halos, but for smaller halos that are barely resolved, the profile improvement is less evident. This is probably due to the fact that in smaller halos the gravitational force is too small and therefore the displacement is not enough. In \cite{vandaalen2015a} it was shown that on scales $2 \lesssim k \lesssim 10 h \mathrm{Mpc^{-1}}$, the power is dominated by the contribution from massive halos $m_{200} \gtrsim 10^{13.5} M_{\odot}$. This explains why the profiles of small halos are not very 
good, while at the same time the power spectrum is matched very well.

\subsection{Substructures}

\begin{figure}[ht]
\includegraphics[width=\textwidth]{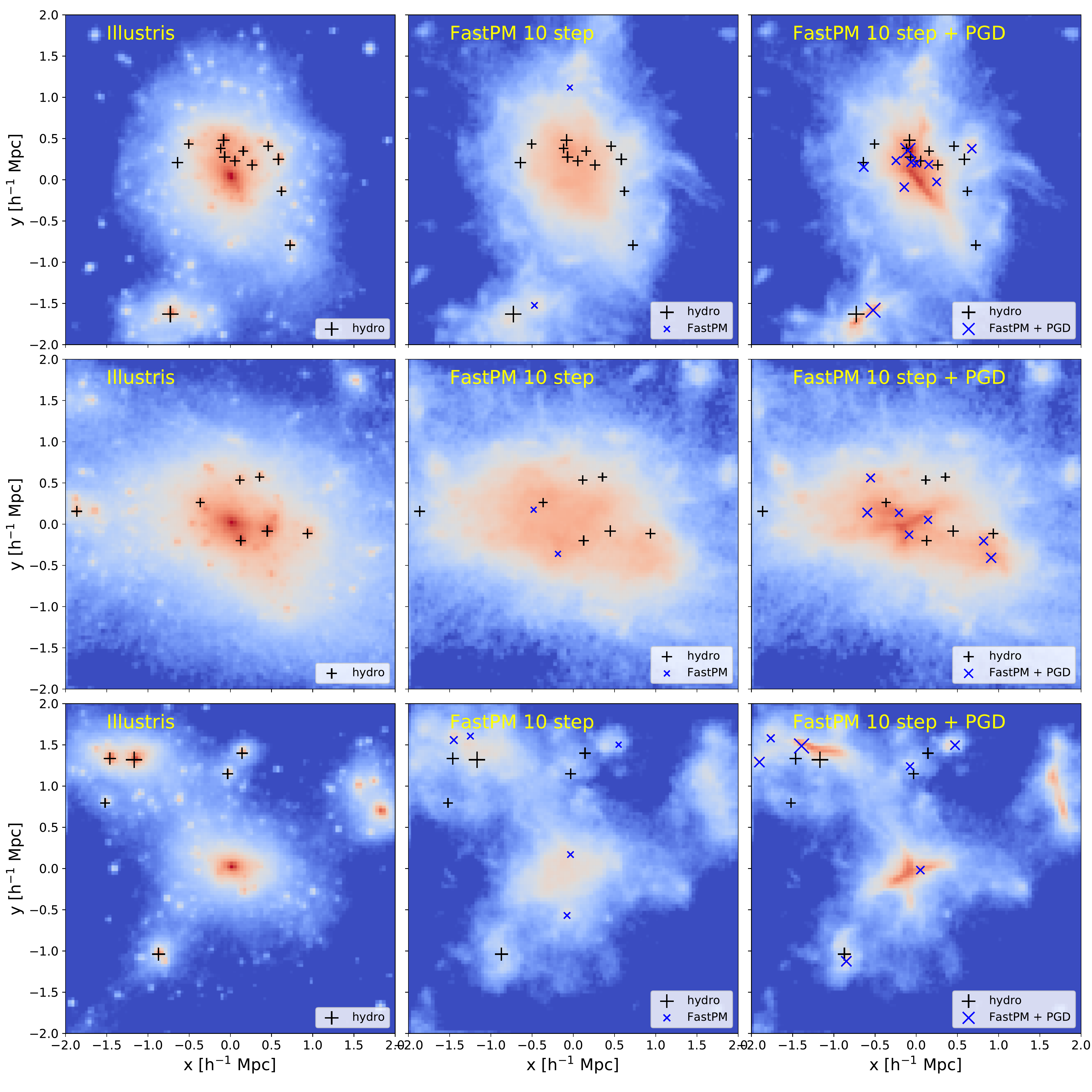}
\caption{Visualizations of some large halos and their satellite sub-halos in Illustris and 10 step \FastPM{}. The satellite sub-halos are indicated with "+" (Illustris) and "$\times$" (\FastPM{}). The PGD model greatly improve the identifications of substructures in 10 step \FastPM{}.}
\label{fig:subhalo10step}
\end{figure}

\begin{figure}[ht]
\includegraphics[width=\textwidth]{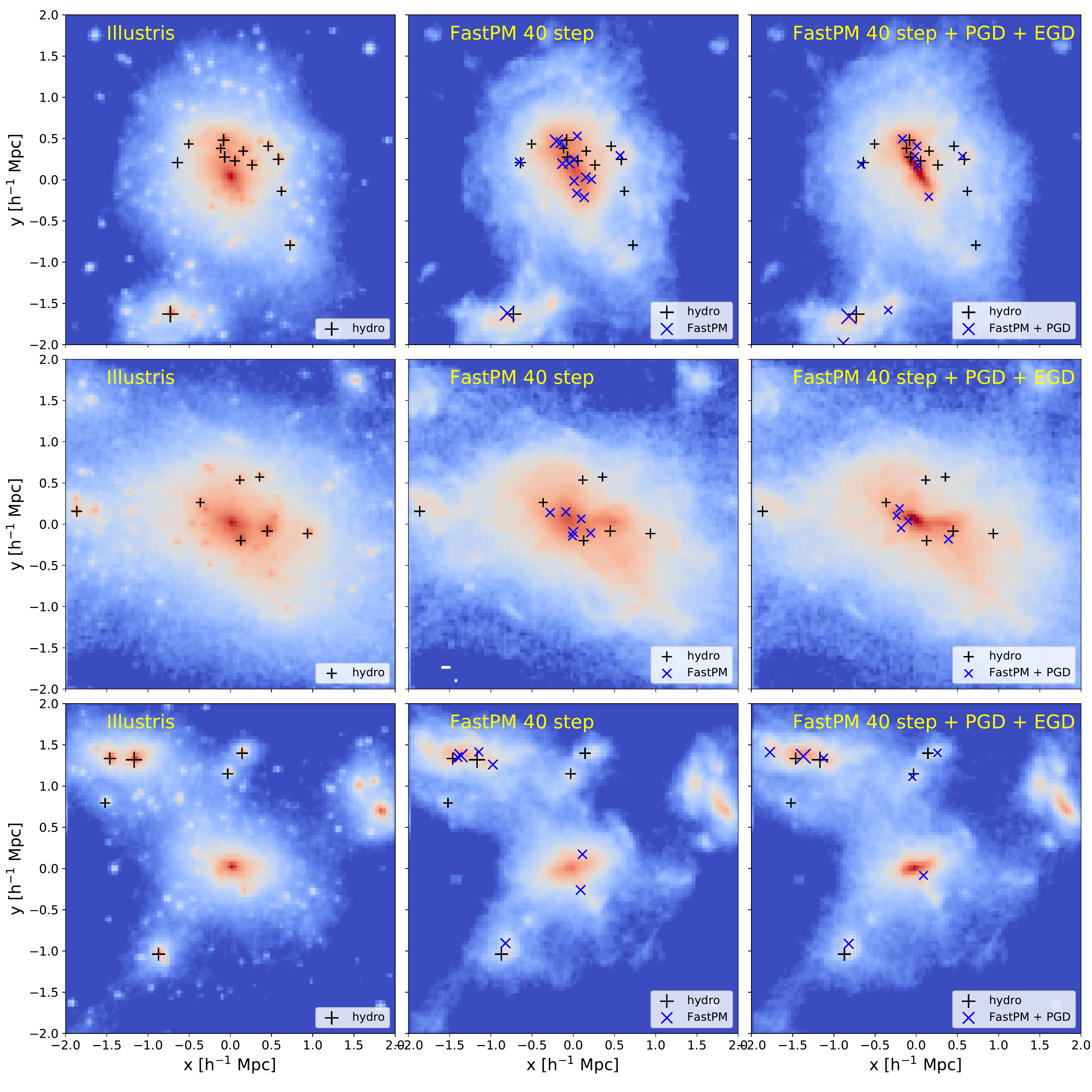}
\caption{Visualizations of some large halos and their satellite sub-halos in Illustris and 40 step \FastPM{}. The satellite sub-halos are indicated with "+" (Illustris) and "$\times$" (\FastPM{}).}
\label{fig:subhalo40step}
\end{figure}

\begin{figure}[ht]
\includegraphics[width=\textwidth]{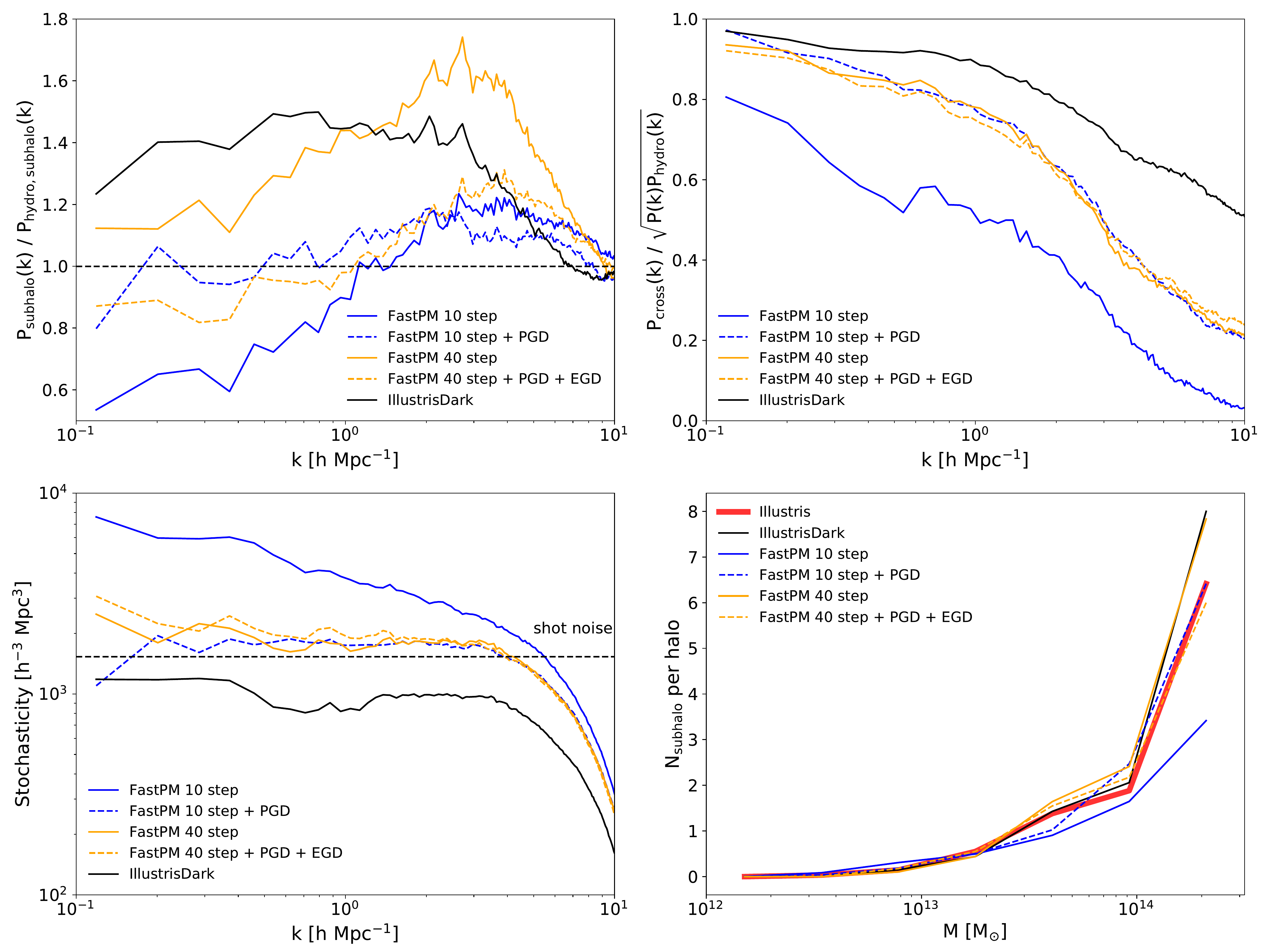}
    \caption{The satellite sub-halo power spectra (upper left), cross correlation coefficients with the reference simulation (upper right), stochasticity (lower left) and the average numbers per halo (lower right). Here we show the results of \FastPM{} 10 step (blue color) and 40 step (orange color) before (straight line) and after (dashed lines) applying our models. We also show the IllustrisDark in black color as a comparison.}
    \label{fig:subhalo-stats}
\end{figure}

Our numerical scheme only affect the matter distribution inside halos, leaving the mass function and clustering of halos intact. We do however expect to see an improvement in the internal structure of halos. In this section we investigate this via the clustering of sub-halos. 

The Illustris-3 simulation provides a sub-halo catalog identified by the SubFind algorithm. Here we choose the satellite sub-halos with $M>10^{12}h^{-1}\rm{M_{\odot}}$ as our reference sub-halo catalog. For \FastPM{} and the calibrated \FastPM{} simulations we use a Friends-of-Friends with a short linking length $l_{\mathrm{FOF}}=0.05$. In practice we find that this choice of linking length works best. Then we remove the central sub-halos, and combine those with distance $r<0.2\hmpc$. Finally the rest are abundance matched with our reference catalog. 

In Figure \ref{fig:subhalo10step} and \ref{fig:subhalo40step} we show an visual inspection of the identified substructures before and after calibration, compared against the Illustris-3 hydrodynamical simulation. We notice that our scheme significantly increases the number of substructures in 10 step \FastPM{}, although they usually tend to be at wrong locations.

In Figure \ref{fig:subhalo-stats} we show the power spectrum, cross correlation coefficient, stochasticity and abundance. We see that our scheme significantly improves all of the statistics. Notably, for 10-step simulation, the cross correlation coefficient improves from 80\% to 95\% at the largest scale, and the stochasticity decreases to the shot noise level. 

\section{Parameter Selection of PGD}
\label{sec:parameter}
\begin{figure}[ht]
\includegraphics[width=\textwidth]{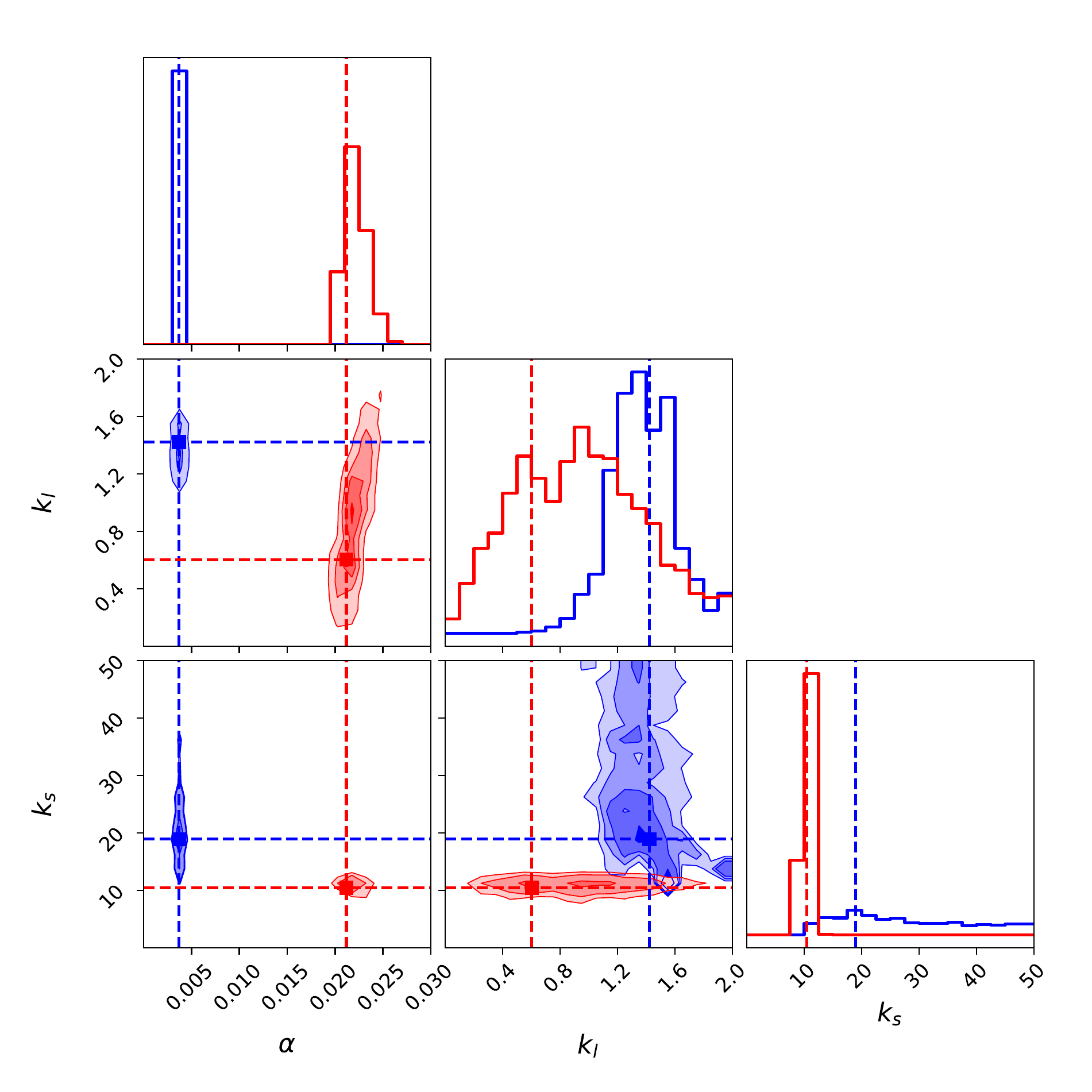}
\caption{The marginalized posterior probability distribution of the 3 parameters in the potential gradient descent model when fitting the power spectra of 10 steps \FastPM{} against Illustris (blue color) and IllustrisDark (red color). The dashed lines show the values of the best fit parameters. Here we set flat priors for these parameters. Because the uncertainties we choose in Section \ref{sec:method} is quite arbitrary, the contours here do not show the true locations of 1 $\sigma$ or 2 $\sigma$. However, they do show us the degeneracy of these parameters (shape of the contours) and the approximated locations of the best fit parameters. Emcee \cite{emcee} is used for sampling.}
\label{fig:posterior}
\end{figure}

\begin{figure}[ht]
\includegraphics[width=\textwidth]{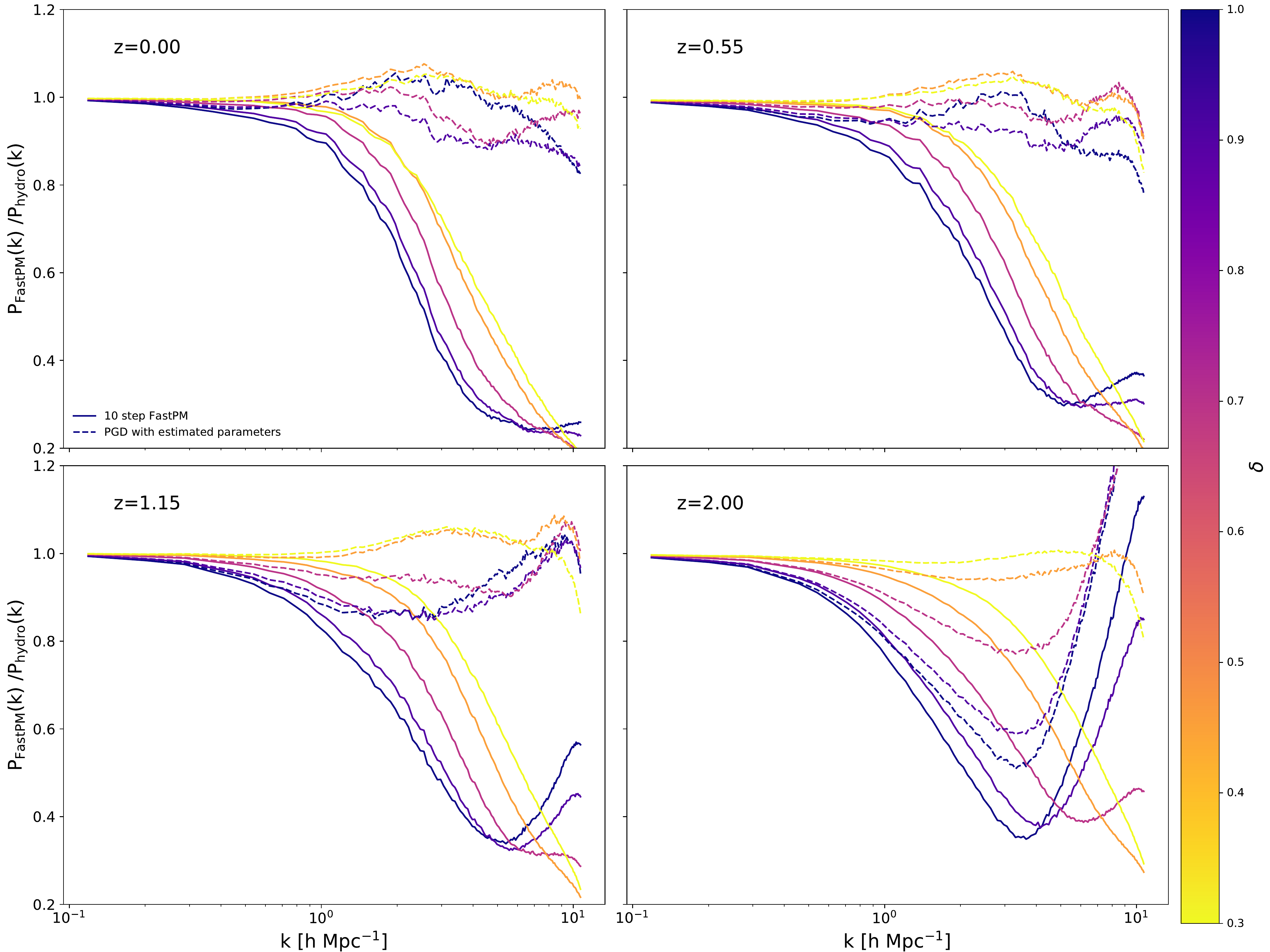}
\caption{The power spectra of \FastPM{} simulation after applying the PGD model. The parameters of PGD model are determined using Equation \ref{eq:paramter1} - Equation \ref{eq:paramter5}. We show the results of different redshifts and of different simulation resolutions ($\delta$ represents the mean separation of particles). At low redshifts, these parameter choices work well for all resolutions; while at high redshifts, low resolutions fail due to shot noise.}
\label{fig:Illustris-power-empiricalrelation}
\end{figure}

The selection of parameters depends on the baryonic physics models, and the parameters of the quasi-nbody simulation, such as mass resolution (typically lower than that of the hydro simulation) and number of time steps. Redshift evolution needs to be calibrated as well. 
On the other hand, we only have a handful of parameters, and their resolution and 
redshift dependencies need to be calibrated only once. 
We note that for 10 step \FastPM{}, PGD alone can provide a significant improvement on the 
power spectrum for a wide range of tests we tried, as seen in Figure \ref{fig:illustris-power-varyalpha}.

We then analyzed the degeneracy of parameters with MCMC of the PGD model in Figure \ref{fig:posterior}. We see that the scheme allows a wide range of $k_l$ or $k_s$ depending on the targeted simulation for calibration. 

It is therefore possible to propose a relatively simple set parameters that cover a set of resolutions. We define the dimensionless resolution parameter $\Delta = \frac{\delta}{1 h^{-1} Mpc}$, where $\delta$ is the mean separation of particles. An approximated empirical relation of the PGD model parameter with $\Delta$ and redshift (as cosmic scale factor $a$) is:
\begin{align}
k_l &= (1.52 - 0.3 \Delta)\ \mathrm{h\ Mpc^{-1}}\label{eq:paramter1}\\
k_s &= (33.4 - 30 \Delta)\ \mathrm{h\ Mpc^{-1}}\label{eq:paramter2}\\
\alpha &= \alpha_0 \cdot a^{\mu}
\label{eq:paramter3}
\end{align}
where
\begin{align}
\alpha_0 &= 0.0061 \Delta^{25} + 0.0051 \Delta^3 + 0.00314\label{eq:paramter4}\\
\mu &= -5.18 \Delta^3 + 11.57 \Delta^2 - 8.58 \Delta + 0.77
\label{eq:paramter5}
\end{align}
The above relation works when $0.3 < \Delta < 1$ and $0 < z < 2$. 

We show the calibrated power spectra using this parameter choice in Figure \ref{fig:Illustris-power-empiricalrelation}. At low redshifts, the power can be calibrated at all resolutions. At high redshift (z=2) the low resolutions do not work well, especially at small scales due to the shot noise. Therefore we recommend using high resolution ($\Delta \lesssim 0.5$) for high redshifts ($z \gtrsim 2$).

We note that the parameter choice presented above is for Illustris baryonic model only. However, it is believed that the Illustris AGN feedback model is too strong and probably unrealistic\citep{genel2014a, haider2016a}. Therefore it is unknown how accurate this parameter choice is.

\section{Conclusions}
\label{sec:conclusion}

In this paper we introduce gradient based method to improve the modeling of matter distribution within halos in low resolution quasi N-body simulations. We train the method on the full N-body  and hydro simulations, with the goal of making the two as close as possible in terms of
summary statistics such as matter power spectrum and halo profiles. We introduce 
two versions of the gradient descent method. The Potential Gradient Descent model drifts the particles along the gradient of modified gravitational potential to help virialize the halos in quasi N-body simulations. In the Enthalpy Gradient Descent model, the particles are moved along the gradient of estimated thermal pressure to model the feedback from AGN and supernovae in a hydro simulation. The latter can also be used by high resolution pure N-body simulations to 
transform them into a hydro simulation. 
We also compare these to the Radial Flow scheme, which naively moves the halo particles along the radial directions to achieve the desired profiles. PGD and EGD are much faster and can 
create some of the substructures, but have very few free parameters, while RF has a lot more 
freedom and can work in all simulations, as long as the halo profiles are given.

We show that all of these models are able to improve the halo profiles and small scale power spectra. The calibrations of PGD and EGD are based on fitting the power spectra. PGD sharpens the halo profiles, especially for massive halos. EGD simulates the AGN feedback by moving the matter to the outskirts of the halos. The effect of cooling and adiabatic contraction is on scales smaller than we are interested in, so here we do not attempt to model these effects. RF calibration is based on calibrating the halo profiles. Both PGD and RF improve the cross correlation coefficients. The PGD model can also improve the subhalo statistics by magnifying the density fluctuation in the halos.

We also present empirical equations of parameter choices for 10 step \FastPM{}, as a function of simulation resolutions and redshifts. This parameter choice gives good results at low redshifts. For high redshifts we recommend using high simulation resolutions to reduce 
the shot noise and improve the small scale power.

PGD and RF can be used to improve the dark matter field in quasi N-body simulations such as \FastPM{} and COLA. Given the reference simulation, the parameters of PGD can be determined by optimizing the power spectrum, and this set of parameters can be used in different realizations and different cosmologies, as long as the simulation resolutions, number of steps and redshifts keep unchanged. We present expressions for these parameters as a function of 
redshift. 
These models will be particularly useful for data analyses where halo internal structures are important, such as weak lensing around galaxies and clusters. PGD will also be useful in galaxy surveys, as it improves the subhalo statistics. EGD can be used to add the baryonic effects to existing dark-matter-only simulations as a first-order approximation.
One limitation of the PGD model is that it does not work that well for small halos. If 
halo profiles at low mass are important, one must increase the force resolution of PM (by going 
to 2 or 3 higher mesh resolution). Alternatively, one can use RF method which can give 
correct average halo profile even at low halo masses. 

There are three free parameters in PGD and two free parameters in EGD. These parameters depend not only on baryonic physics models, but also on simulation resolutions, redshifts, and the number of steps. These parameters cannot be derived from the first principles. To achieve the best results, they need to be optimized for different situations. This can be viewed as a 
positive aspect of this approach: we do not fully understand the physical proceses that 
govern feedback models and their impact on the halo mass profiles, so these astrophysical 
uncertainties must be modeled with free parameters. Our approach manages to compress 
the number of free parameters down to a few only, so it 
can be used as a useful parametrization of our astrophysical ignorance that needs to 
be marginalized over. In this sense one can argue that high resolution N-body simulations 
are no better than low resolution simulations: they are both missing baryonic effects, 
and if these effects can be incorporated with a few unknown nuisance parameters into either 
scheme with equal results then 
the  advantages of the high resolution N-body simulations are eliminated.

In the future we plan to incorporate this scheme into FastPM and its gradient 
of final density field with respect to initial density modes, which are needed for the reconstruction of initial conditions \cite{seljak2017a}. This will be particularly important if the data for 
reconstruction include weak lensing, which can resolve halo mass profiles, at least statistically. 
We also plan to investigate ways of embedding the scheme directly into the simulation as additional viscous drifting that bypasses the momentum, as a way to alleviate the redshift dependency of the parameters, and as a way to produce further enhanced weak lensing maps.
Finally, it is well known that the AGN feedback model of Illustris is too strong, so in some 
sense the parameters we determine span teh maximal range of baryonic effects. 
In the future 
we plan to test the method on several additional baryonic feedback simulations, 
to verify and if needed expand the parameter space of baryonic parameters, and determine 
their most likely values. 

\textbf{Acknowledgements}
The majority of the computation were performed on NERSC computing facilities Edison and Cori, billed under the cosmosim and m3058 repository. We acknowledge support of NASA grant NNX15AL17G. 
We thank Simeon Bird for kindly providing the linear power spectrum and random seed of Illustris simulation.

\appendix
\section{Simulation data sets}
\label{sec:simulation}
\subsection{Illustris-3/IllustrisDark-3}

Illustris \citep{vogelsberger2014a, vogelsberger2014b,genel2014a} is a series of cosmological hydrodynamic simulations, carried out with the moving-mesh code AREPO \citep{springel2010a}. Each simulation evolved a periodic volume 106.5 Mpc on a side, over the redshift range $z = 127$ to the present in a $\mathrm{\Lambda CDM}$ cosmology($\Omega_M = 0.2726$, $\Omega_b = 0.0456$, $H_0=70.4 {\rm km\ s^{−1}\ Mpc^{-1}}$, $n_s= 0.963$, $\sigma_8 = 0.809$). Illustris follows the evolution of the dark matter, cosmic gas, stars and supermassive black holes, with a full set of physical models including primordial and metal-line gas cooling, star formation and evolution, gas recycling, chemical enrichment, supernova feedback and AGN feedback (for more details see \citep{vogelsberger2013a, torrey2014a}).

Illustris has three runs (Illustris-1,2,3) at different resolutions. Since the scale we are interested in this study is larger than $k = 10 hMpc^{-1}$, we focus on Illustris-3, which has a mass resolution $m_{\mathrm{DM}}=2.8\times10^8 h^{-1} M_{\odot}$, $\bar m_{\mathrm{baryon}}=0.57\times10^7 h^{-1} M_{\odot}$ and a force resolution $\epsilon_{\mathrm{DM}}=5.68\mathrm{kpc}$, $\epsilon_{\mathrm{baryon}}=2.84\mathrm{kpc}$. As a comparison, we also make use of Illustris-3-Dark, a dark-matter-only analog of Illustris-3.


\subsection{FastPM}
\FastPM{} is a quasi particle-mesh (PM) N-body solver, in which the drift and kick factors are modified following the Zel'dovich equation of motion so that the correct linear theory growth at large scale can be produced at a limited number of steps. 
We generate the initial condition with the same random seed and linear power spectrum as 
Illustris, starting at $z=9$ and using the second order Lagrangian perturbation theory. 

\section{Cost Function and Choice Covariance}
\label{sec:method}

The calibrations of potential gradient descent and enthalpy gradient descent models are based on minimizing the discrepancies of the power spectra. The covariance matrix of power spectrum in the fully nonlinear scale is unknown. To avoid this complication, we assume a simple Gaussian likelihood that weights different scales equally. 
\be
\label{eq:likelihood}
p(P_{\mathsf{ref}}(k)|\theta) = \prod_{k<10 \mathsf{h\ Mpc^{-1}}}\frac{1}{\sqrt{2\pi \sigma_k^2}}\exp[\frac{(P_{\mathsf{ref}}(k) - P_{\mathsf{calib}}(k))^2}{2\sigma_k^2}] .
\ee
where $\theta$ represents the parameters in our models, and $\sigma_k$ is the error of $P_{\mathsf{ref}}(k)$. One natural choice for $\sigma_k$ is $\sigma_k = \sqrt{\frac{2}{N}} P_{\mathsf{ref}}(k)$, where $N$ is the number of k modes in this bin. However, in practice we find that this choice often overemphasizes the small scale power, as on small scale $\sigma_k$ is quite small due to the large number of k modes. As a result, on scale $k \approx 3 \mathsf{h\ Mpc^{-1}}$ the fitting is quite poor, even though on smaller scale the power matches well. We argue that in observation the error of small scale power is often dominates by systematic error, and is much larger than $\sqrt{\frac{2}{N}} P_{\mathsf{ref}}(k)$. We try to avoid this by choosing $\sigma_k = 0.1 P_{\mathsf{ref}}(k)$. The factor $0.1$ here does not change the best fit parameters. We find that this $\sigma_k$ choice works well, as shown in the paper.

We also attempt to maximize the cross correlation coefficient, which gives a similar improvement in the correlation coefficient but a drastically different small scale power spectrum. 

A third cost function we attempted is directly matching the density field in configuration space by minimizing the residual,
\be
y = \sum_{x_i}(\delta_{\mathsf{calib}}(x_i) - \delta_{\mathsf{ref}}(x_i))^2.
\ee
We find that the configuration residual down-weights the large scale power (due to 
fewer modes), preferring parameters that produces incorrect large scale power.

\bibliographystyle{revtex}
\bibliography{reference}

\end{document}